\documentclass[aip, amsmath,amssymb, reprint,]{revtex4-1}%
\usepackage{graphicx}
\usepackage{bm}
\usepackage{epstopdf}
\usepackage{sidecap}
\usepackage{amsmath}
\usepackage{amsfonts}
\usepackage{amssymb}%
\providecommand{\U}[1]{\protect\rule{.1in}{.1in}}
\begin{document}
\title{An ultra-thin waveguide twist constructed using fish-scale metallic wires}

\author{Jin Han}
\affiliation{Department of Physics, Tongji University, Shanghai 200092, People's Republic of China}
\affiliation{Shanghai Key Laboratory of Special Artifical Microstructure Materials and Technology, Shanghai, People's Republic of China}
\author{Hongqiang Li}
\email{hqlee@tongji.edu.cn}
\affiliation{Department of Physics,
Tongji University, Shanghai 200092, People's Republic of China}
\affiliation{Shanghai Key Laboratory of Special Artifical
Microstructure Materials and Technology, Shanghai, People's Republic
of China}
\author{Yuancheng Fan}
\affiliation{Department of Physics, Tongji University, Shanghai 200092, People's Republic of China}
\affiliation{Shanghai Key Laboratory of Special Artifical Microstructure Materials and Technology, Shanghai, People's Republic of China}
\author{Zeyong Wei}
\affiliation{Department of Physics, Tongji University, Shanghai 200092, People's Republic of China}
\affiliation{Shanghai Key Laboratory of Special Artifical Microstructure Materials and Technology, Shanghai, People's Republic of China}
\author{Chao Wu}
\affiliation{Department of Physics, Tongji University, Shanghai 200092, People's Republic of China}
\affiliation{Shanghai Key Laboratory of Special Artifical Microstructure Materials and Technology, Shanghai, People's Republic of China}
\author{Yang Cao}
\affiliation{Department of Physics, Tongji University, Shanghai 200092, People's Republic of China}
\affiliation{Shanghai Key Laboratory of Special Artifical Microstructure Materials and Technology, Shanghai, People's Republic of China}
\author{Xing Yu}
\affiliation{Department of Physics, Tongji University, Shanghai
200092, People's Republic of China} \affiliation{Shanghai Key
Laboratory of Special Artifical Microstructure Materials and
Technology, Shanghai, People's Republic of China}
\author{Fang Li}
\affiliation{Department of Physics, Tongji University, Shanghai
200092, People's Republic of China} \affiliation{Shanghai Key
Laboratory of Special Artifical Microstructure Materials and
Technology, Shanghai, People's Republic of China}
\author{Zhanshan Wang}
\affiliation{Department of Physics, Tongji University, Shanghai
200092, People's Republic of China} \affiliation{Shanghai Key
Laboratory of Special Artifical Microstructure Materials and
Technology, Shanghai, People's Republic of China}

\begin{abstract}
This study theoretically and experimentally investigates the
transmission properties of a metamaterial slab comprised of two
layers of metallic fish-scale structure arrays and a sandwiched
dielectric layer. Calculations show that the asymmetric transmission
can be tuned by varying the slab thickness, due to evanescent
interlayer coupling. The spatial evolution of the local field inside
the structure indicates that the slab functions as a perfect
polarization transformer at certain frequencies in the manner of a
waveguide twist. Measured transmission spectra are in good agreement
with calculated results when material dissipation is considered.

\end{abstract}
\maketitle Recently, considerable interests in the asymmetric
transmission (AT) of polarized light has been revived in the context
of planar chiral metamaterials\cite{1,2,3,4,5,6,7,8}.This
phenomenon, arising from the vector nature of electromagnetic waves,
still satisfies the de Hoop reciprocity as revealed by Jones and
Muller matrix formulation\cite{9}. That is, in the transfer-matrix
representation of an optical system, the matrix for reverse
transmission is the transpose of that of forward transmission. The
system will give rise to asymmetric transmission of polarized light
provided that the system is chiral and anistropic\cite{7}. The
metamaterial approach\cite{1,2,3,4,5,6,7,8,9} enables this
functionality to be implemented at subwavelength thicknesses. These
findings constitute an alternative solution for one-way
electromagnetic isolator and related devices to the scheme relying
on nonreciprocal ingredients such as gyrotropy\cite{10}, or
nonlinearity\cite{11}et al.

It is notable that the coupling among the metallic building blocks
or meta-atoms is instrumental for the exotic properties of the
metamaterial. A microscopic investigation is helpful for deeper
insight into the underlying physics as well as for practical
applications. Fish-scale metallic wire\cite{12}, taken as a typical
example of a planar chiral metamaterial, has been investigated
extensively as an optical magnetic mirror\cite{13} and local field
concentrator\cite{14}. In this paper, we propose a device for
asymmetric transmission that is composed of two layers of fish-scale
structure metallic wires in an orthogonal arrangement and a
sandwiched dielectric spacer layer. Calculation of the local field
distribution reveals that the thin slab functions as a waveguide
twist\cite{15,16}, giving rise to nearly perfect asymmetric
transmission. This physical picture unifies mode matching at the
entrance interface and polarization transformation inside the slab
to explain the formation of AT and its dependence on slab thickness.

\begin{figure}[pb]
\includegraphics[width=8.6cm
]{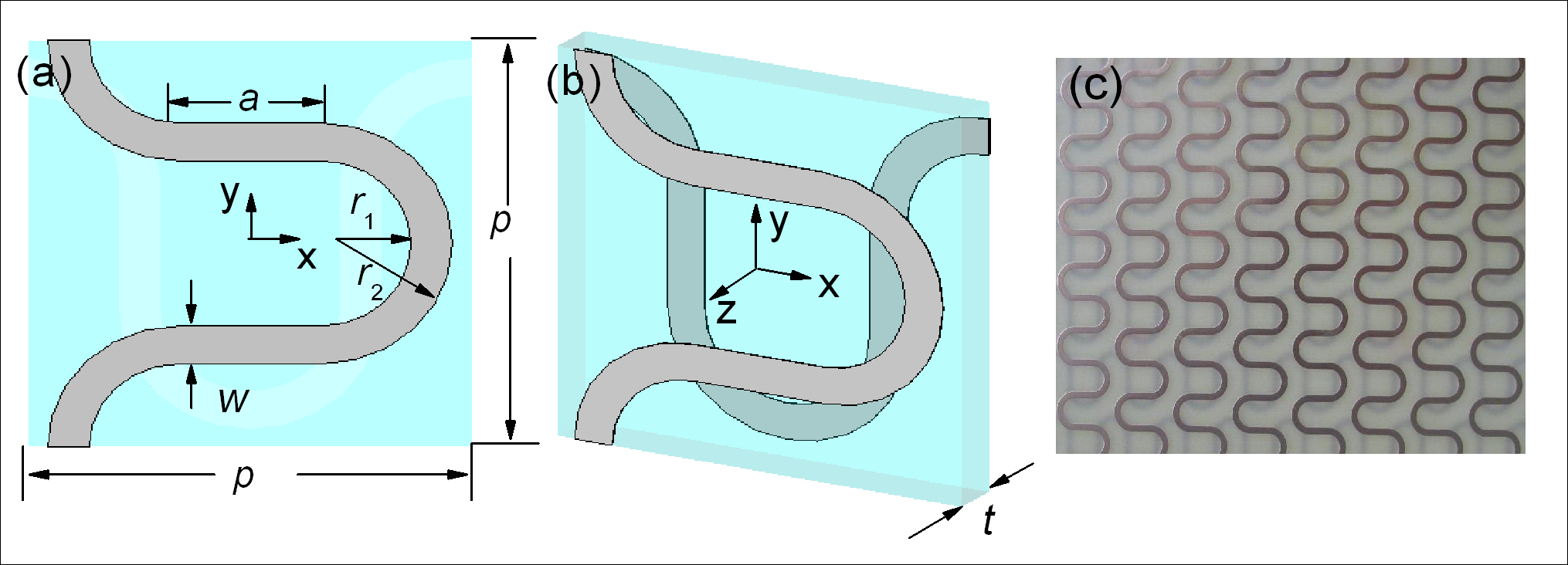}\caption{a schematic of (a)the front view (in $x-y$ plane)
of the sample slab, (b)the stereogram of a unit cell, (c)a photo of
our sample.
The thickness of the dielectric layer is $t=$1.8mm.}%
\end{figure}

\begin{figure}[pb]
\includegraphics[width=8.6cm
]{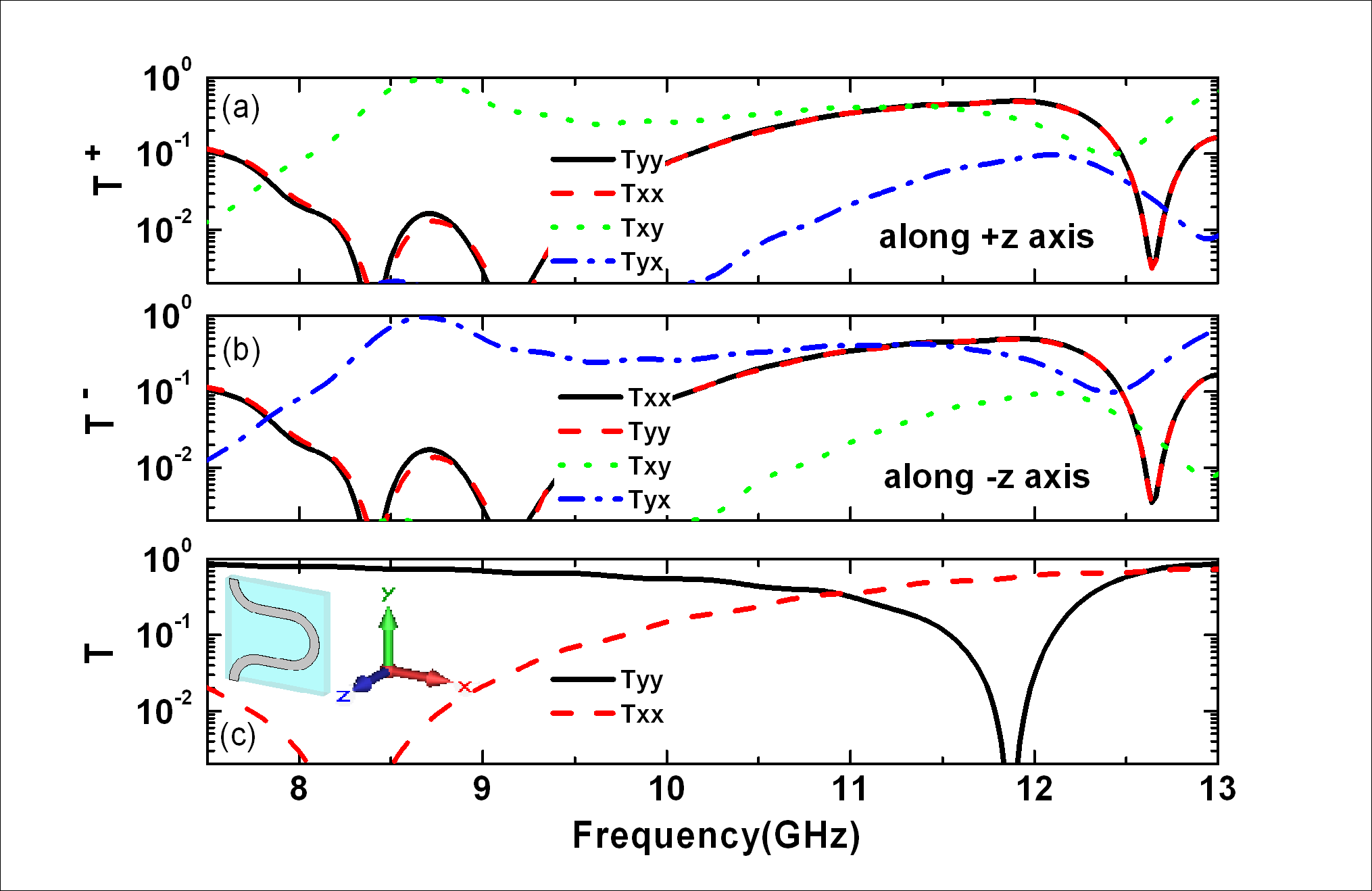}\caption{Calculated transmission spectra for $x$-polarized
$(\vec{E}\parallel\vec{x})$ and $y$ polarized
$(\vec{E}\parallel\vec{y})$ incidence of our sample slab (a)along
$+z$ direction and (b) along $-z$ direction. (c)a monolayer of
fish-scale structure metallic arrays with $y$-oriented wires shown in Fig. 1(a).}%
\end{figure}

A unit cell of our model slab, shown in front view and stereogram in
Fig. 1(a) and Fig. 1(b), is composed of circularly meandering and
straight metallic strips. The metallic patterns on the front and
back surfaces of the dielectric layer are the same, but are rotated
relative to each other at an angle of $90^{0}$ about the $z$ axis.
The geometric parameters marked in Fig. 1(a) are the width $w=1.4$mm
of the metallic strips, the radii $r_{1}=3.05$mm, $r_{2}=4.45$mm of
the inner and outer circles, the length $a=4.8$mm of the straight
metallic strip, and period $p=15$mm of a square unit cell. The
substrate has a thickness of $t=1.8$mm. Fig. 1(c) shows a photo of
our sample with the same geometric parameters as the theoretical
model in Fig. 1(a). The metallic fish-scale patterns are fabricated
and deposited on the $35\mu m$-thick copper foils on both sides of
an FR4 dielectric substrate. The lateral size of the entire sample
slab is $570\times435$mm.

\begin{figure}[ptb]
\centerline{\includegraphics[width=8.6cm]{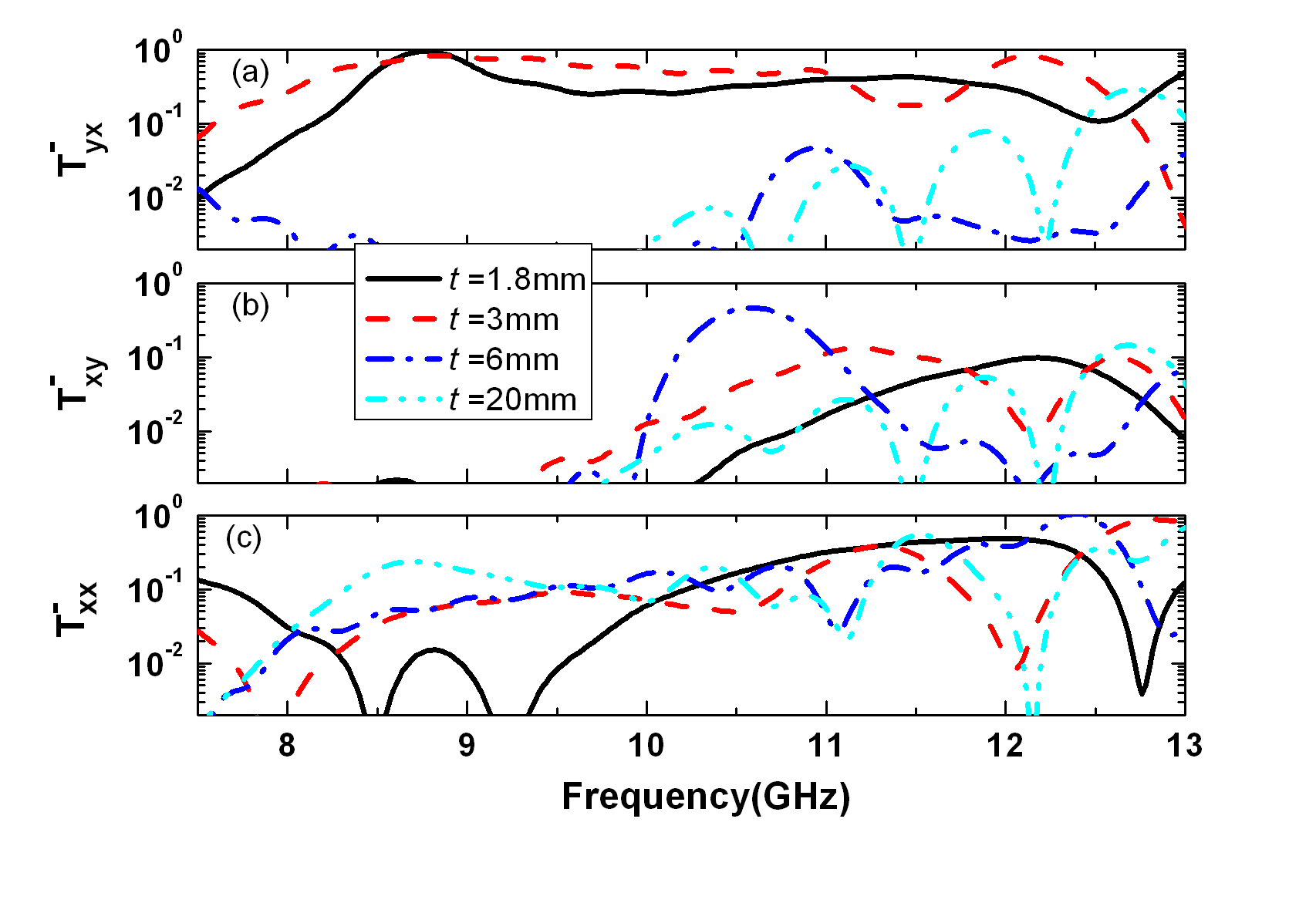}} \caption{ The
spectra of backward transmission for different slab thicknesses
$t$.}\label{fig2}
\end{figure}

\begin{figure*}[pth]
\centerline{\includegraphics[width=15cm]{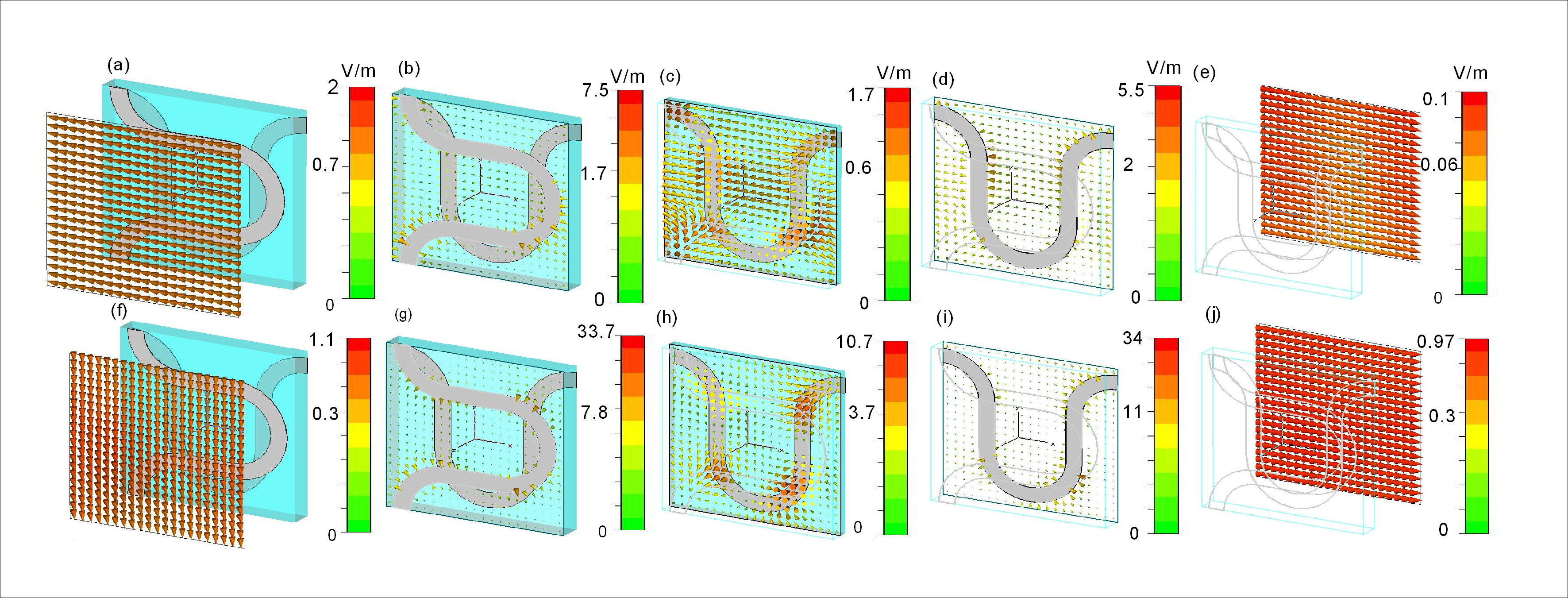}} \caption{ Snapshots
of the electric field strength at different $z$ coordinates for
$x$-polarized ((a)-(e)) and $y$-polarized ((f)-(j)) incident plane
waves along the $-z$ direction. (a) and (f): at $z=7.4$mm of the
incoming regime; (b) and (g): at $z=0$ of the front
surface(entrance); (c) and (h) at $z=-1.1$mm within the dielectric
layer; (d) and (i) at $z=-1.8$mm of the backside surface; (e) and
(j): at $z=-8.4$mm of the outgoing regime. The local field strength,
as illustrated by the color bar, is normalized to that of the
incident wave.}\label{fig2}
\end{figure*}

We perform finite-difference-in-time-domain (FDTD) numerical
simulations to calculate the transmission spectra of our fish-scale
model for $x$-polarized and $y$-polarized incident waves along the
$+z$ and $-z$ directions. For simplicity, metals are assumed to be
perfect electric conductors (PECs). The dielectric substrate is
assumed to be lossless, with a permittivity
$\epsilon_{r}\approx4.1$, the same as that of the FR4 substrate. The
transmission coefficient $T_{ij}^{d}$  is defined in terms of the
intensity of the electric field, with the first and the second
subscripts denoting the polarized states of the incident wave and
transmitted wave components, respectively, and the superscript
$d=+/-$ referring to forward$(+)$ or backward$(-)$ transmission. For
example, $T_{xy}^{+}$ denotes the normalized intensity of the
$y$-polarized transmitted wave component$(\vec{E}\parallel\vec{y})$
with respect to the $x$-polarized$(\vec{E}\parallel\vec{x})$
incident wave along the forward direction ($k_{z}$$>$$0$ and
$k_{\parallel}$$=$$0$), and the total transmissivity (for all
polarizations) is given by $\mid T_{xx}^{+}+T_{xy}^{+}\mid$.

Fig. 2(a) and Fig. 2(b) presents the forward and backward
transmission spectra $T^{+}$ and $T^{-}$ of our model slab
calculated by the FDTD algorithm. We see, both in Fig. 2(a) and Fig.
2(b) that $T_{xx}$ is nearly superposed onto $T_{yy}$. This is
because the structure, taken as a whole, is essentially isotropic,
due to the orthogonal alignment of the fish-scale patterns on both
sides of the dielectric layer. However, a salient feature in Fig.
2(a) and Fig. 2(b) is that $T_{xy}$ is markedly different from
$T_{yx}$ at all frequencies. $T_{xy}$ and $T_{yx}$ are interchanged
when the incident direction (wavevector $\vec{k}$) is reversed,
giving rise to a clear signature of asymmetric transmission. A
maximum of 97\% transmission is reached at $8.7$GHz in the spectrum
of $T_{xy}^{+}$; meanwhile, both $T_{yx}^{+}$ and $T_{xx}^{+}$ are
below 1\% transmission near $8.7$GHz.

The asymmetric property described above has close connections with
the transmission spectra of a monolayer of fish-scale metallic
wires. As shown in Fig. 2(c), the monolayer with $y$-oriented wires
is transparent to the $y$-polarized incident wave
($T_{yy}$$>$$60\%$) around $8.7$GHz but opaque to the $x$-polarized
incident wave ($T_{xx}$$<$$10\%$) at the same frequencies. This is
because there exists a local mode owned by the $x/y$-oriented
fish-scale structure that only couples to $x/y$-polarized plane
waves near the frequency $8.7$GHz\cite{12}. In the case presented in
Fig. 2(a), an $x$-polarized incident wave at approximately $8.7$GHz
in the forward direction will meet the $x$-oriented fish-scale
patterns first, so that it can propagate inside the slab and be
transformed into $y$-polarization afterwards. In contrast, an
$y$-polarized incidence at the same frequency can not couple with
the local mode based on the metallic layer at the entrance. Similar
rules also apply to a backward incident wave along the $-z$
direction, as seen by interchanging $x$ and $y$ subscripts (see Fig.
2(b)). However, asymmetric transmission with high bandwidth, as
shown in Fig. 2(a) and Fig. 2(b), can not be deduced from the
transmission properties of a monolayer of fish-scale metallic
patterns (Fig. 2(c)). The coupling between the metallic layers must
be considered for a complete description of theses phenomena.

Fig. 3 presents the calculated transmission spectra $T_{yx}^{-}$,
$T_{xy}^{-}$ and $T_{xx}^{-}$ with respect to the different
thicknesses $t=1.8$mm, $3$mm, $6$mm and $20$mm of the dielectric
layer. We see from Fig. 3(a) that, $T_{yx}^{-}$ varies rapidly in
lineshape as a function of slab thickness $t$. When $t$ is small,
both $T_{xx}^{-}$ and $T_{xy}^{-}$ are suppressed very well at a
level of $10^{-2}-10^{-3}$, within the $7.5$$-$$10$GHz range, giving
rise to an AT functionality with good performance. At $t=6$mm, the
AT below $10$GHz disappears. However, AT is observed along the
reverse direction near $10.5$GHz where $T_{xy}^{-}$ reaches its
maximum of 0.46, and $|T_{yx}^{-}+T_{xx}^{-}|$ $\approx$ $0.027$ is
still very small. This means that the asymmetric transmission can be
tuned to a different frequency even along the reversed direction, by
changing the thickness $t$. For much larger values of $t$, such as
$t=20$mm, the evanescent interlayer coupling is negligible and the
AT effect disappears at all frequencies.

\begin{figure}[ptb]
\includegraphics[width=8.6cm
]{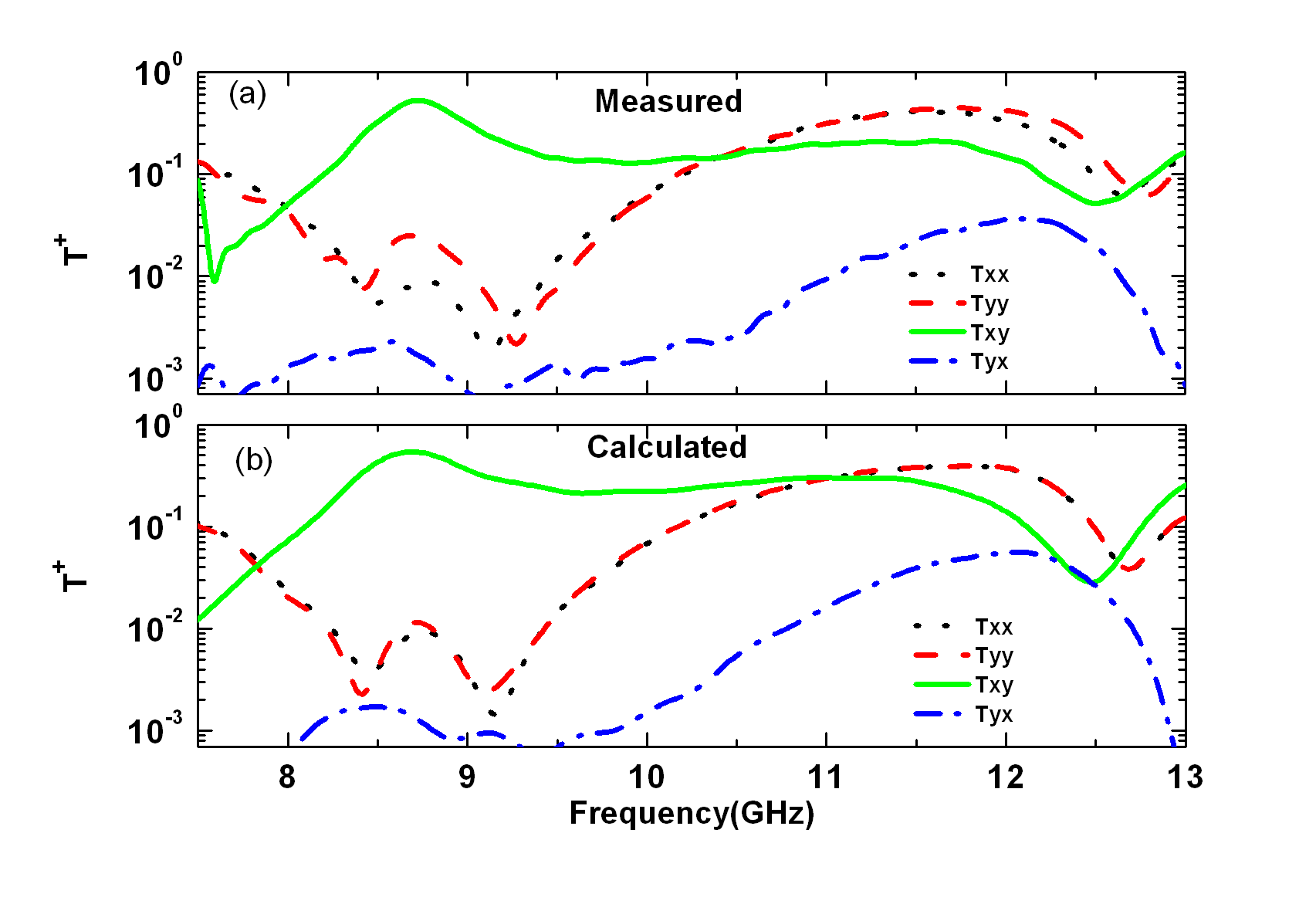}\caption{(a) measured and (b) calculated forward transmission of our sample slab.}%
\end{figure}

A direct way to witness the formation of asymmetric transmission is
to observe the evolution of wave fields propagating inside the model
slab. The snapshots in Fig. 4 calculated using the FDTD algorithm
show that, for backward transmission at $8.7$GHz, the outgoing waves
are always $x$-polarized (see Fig. 4(e), (j)) even when the incident
wave is $y$-polarized (Fig. 4(f)). This is because only the
$x$-polarized wave component at $8.7$GHz is allowed to pass through
the $x$-oriented fish-scale metallic arrays at the outgoing surface.
We see from Fig. 4(b)-(d) and Fig. 4(g)-(i) that the field patterns
are twisted inside the slab as a consequence of interlayer coupling.
We also see that the field strength inside the slab remains high
(Fig. 4(c)), even when the incident wave is blocked at the entrance
surface. These are the features of evanescent coupling between
metallic layers. The picture demonstrates that our model slab can
function as an ultra-thin polarization transformer just as a
waveguide twist with $90^{0}$ orientation would function.

The transmission spectra through our sample (see Fig. 1(c)) were
measured with a vector network analyzer (Agilent 8722ES) in an
anechoic chamber. The spectra for forward transmission are shown in
Fig. 5(a). A peak value of $0.53$ was measured at $8.7$GHz in the
spectrum of $T_{xy}$. Measurements are quantitatively verified by
calculations (see Fig. 5(b)) adopting a complex permittivity of
$\epsilon_{r}\approx4.1+0.1i$ for the FR4 substrate. Almost half of
the wave energy is removed by the absorption channel (not shown) due
to a strong enhancement of the local field inside the slab, which
maximizes dielectric dissipation. The reflection at $8.7$GHz is
still very small.

In conclusion, we provide a picture of waveguide twist to explain
the AT effect of our model slab, which contains two layers of
fish-scale metallic wires. Calculations on the AT with respect to
different slab thicknesses and the spatial evolution of the local
field verifies that near-field interlayer coupling enables the
polarization transformation of the outgoing waves, whereas the
linearly polarized incident wave can only propagate into the slab in
one direction, giving rise to one-way asymmetric transmission. Our
findings are beneficial for the design of optical isolators based on
planar chiral metamaterials. This work was supported by NSFC (No.
10974144, 60674778), the National 863 Program of China (No.
2006AA03Z407), NCET (07-0621), STCSM, and SHEDF (No. 06SG24).


\begin{thebibliography}{99}%
\bibitem {1}V. A. Fedotov, A. S. Schwanecke, N. I. Zheludev, V. V. Khardikov, and S. L. Prosvirnin, Nano Lett. \textbf{7,} 1996 (2007).
\bibitem {2}E. Plum, V. A. Fedotov, and N. I. Zheludev, Appl. Phys. Lett. \textbf{94,} 131901 (2009).
\bibitem {3}S. V. Zhukovsky, A. V. Novitsky, and V. M. Galynsky, Opt.Lett. \textbf{34,} 1988 (2009).
\bibitem {4}E. Plum, V. A. Fedotov, and N. I. Zheludev, arXiv.org.1006.0870 (2010).
\bibitem {5}R. Singh \emph{et al}.,  Phys. Rev. B. \textbf{80,} 153104 (2009).
\bibitem {6}S. I. Maslovski, D. K. Morits, and S. A. Tretyakov, J. Opt. A: Pure Appl. Opt. \textbf{11,} 074004 (2009).
\bibitem {7}V. A. Fedotov, P. L. Mladyonov, S. L. Prosvirnin, A. V.Rogacheva, Y. Chen, and N. I. Zheludev, Phys. Rev.Lett. \textbf{97,} 167401 (2006).
\bibitem {8}C. Menzel, C. Helgert, C. Rockstuhl, E.-B. Kley, A. Tunnermann, T. Pertsch, and F. Lederer, Phys.Rev.Lett. \textbf{104,} 253902 (2010).
\bibitem {9}R. J. Potton, Rep. Prog. Phys.\textbf{67,} 717 (2004).
\bibitem {10}V. M. Agranowitz  and V. L. Ginzburg, Spatial Dispersion in Crystal Optics and the Theory of Excitons (Wiley:London)(1966)
\bibitem {11}Y. Fan, J. Han, Z. Wei, C. Wu, Y. Cao, X. Yu, and H. Li, arXiv.org.1012.2252 (2010).
\bibitem {12}V. A. Fedotov, P. L. Mladyonov, S. L. Prosvirnin, and N. I. Zheludev, Phys. Rev. E. \textbf{72,} 056613 (2005).
\bibitem {13}A. S. Schwanecke \emph{et al}., J. Opt. A: Pure Appl. Opt. \textbf{9,} L1-L2 (2007).
\bibitem {14}T. S. Kao \emph{et al}., Appl. Phys. Lett. \textbf{96,} 041103 (2010).
\bibitem {15}H. A. Wheeler, and H. Schwiebert, IRE Trans. Microwave Theory Tech. \textbf{MTT-3,} 44(1955).
\bibitem {16}B. C.DeLoach,IRE Trans. Microwave Theory Tech. \textbf{MTT-9,}130 (1961).

\end{thebibliography}
\end{document}